\documentclass[
reprint,
showpacs,
nofootinbib,
amsmath,amssymb,
aps,
pra,
floatfix,
]{revtex4-1}

\usepackage{graphicx}
\usepackage{dcolumn}
\usepackage{bm}
\usepackage[colorlinks=true, citecolor=blue, urlcolor=blue ]{hyperref}
\usepackage{pxfonts,txfonts}

\newcommand{\bq}{\begin{equation}}
\newcommand{\eq}{\end{equation}}
\newcommand{\bqs}{\begin{equation*}}
\newcommand{\eqs}{\end{equation*}}
\newcommand{\ba}{\begin{array}}
\newcommand{\ea}{\end{array}}
\newcommand{\bas}{\begin{array*}}
\newcommand{\eas}{\end{array*}}
\newcommand{\bqa}{\begin{eqnarray}}
\newcommand{\eqa}{\end{eqnarray}}
\newcommand{\bqas}{\begin{eqnarray*}}
\newcommand{\eqas}{\end{eqnarray*}}

\newcommand{\etal}{\textit{et al.} }

\newcommand{\D}{\mathcal{D}}

\newcommand{\qed}{\nobreak \ifvmode \relax \else
\ifdim\lastskip<1.5em \hskip-\lastskip
\hskip1.5em plus0em minus0.5em \fi \nobreak
\vrule height0.3em width0.5em depth0.25em\fi}

\begin{document}


\title{Comment on ``Sudden change in quantum discord accompanying the transition from bound to free entanglement''}
\author{Swapan Rana}
\email{swapanqic@gmail.com}
\author{Preeti Parashar}
\email{parashar@isical.ac.in}
\affiliation{Physics and Applied Mathematics Unit, Indian Statistical Institute, 203 B T Road, Kolkata, India}

\begin{abstract} In a recent article [\href{http://dx.doi.org/10.1103/PhysRevA.87.022340}{Phys. Rev. A {\bf 87}, 022340 (2013)}], Yan \etal  have  studied geometric discord for a well known class of bound entangled states. Based on their calculation, they claim ``It is found that there exists a nondynamic sudden change in quantum discord'' for these states. In this Comment, we criticize their work by pointing out that what has been calculated is actually a bound, and not the exact value of discord.  Since, generally, it is not possible to infer the exact value (or property) of a quantity just from its bound, we can not conclude about (exact) discord and its change. Thus, the above-mentioned conclusion can not be drawn from the calculations of the paper. 

\end{abstract}

\maketitle


In Ref.~\cite{YanLiuCheePRA13}, the authors have studied geometric discord for a well-known class of $3\otimes 3$ bound entangled states [see Eq.~(16) therein for the state]. The main result of their paper is ``there exists a nondynamic sudden change in quantum discord'' for these states. The purpose of this Comment is to criticize the paper by pointing out that this conclusion can never follow from their calculations.

Before elaborating on the major drawbacks, let us mention a small typographical error. The eigenvalues calculated are of $CC^t$ and not of $C$. Also the authors did not write explicitly the expression of the sought bound.

Let us now clarify a serious misconception in Ref.~\cite{YanLiuCheePRA13}. The authors claimed their result about \textit{sudden change in quantum discord} from Fig.~1, which is supposed to depict quantum discord with respect to $\alpha$ (indeed the vertical axis is explicitly labeled as ``quantum discord"). However, we note that all that has been calculated is only a bound and not the exact value of discord. Since it is impossible to infer the exact value of a quantity just from its bound, the calculated bound can not be used to draw conclusion about the exact discord and, hence, its change. Indeed the correct way to evaluate discord is to calculate this bound and then find some optimal measurements (or equivalently \textit{classical-quantum} states) saturating the bound. However, it may happen that there is no such measurement agreeing with the bound, thereby, the bound is strictly smaller than the exact discord. This, indeed happens for most generic $3\otimes 3$ states, because the bound (even the sharpest one) is exact only for $2\otimes n$ states. In particular, it is well known that the bound used in Ref.~\cite{YanLiuCheePRA13} is quite a rough one (see the following paragraph for details). So, unless verified with some measurements, it is impossible to say just from the calculated bound that the graph of (exact) discord would not be a straight line, instead of the curve shown. Thus, the author's conclusion about sudden change of (exact) discord does not follow from the calculations presented in their paper. For the same reason, Fig.~1 does not depict quantum discord, but only its bound.

 The use of formula in Eq.~(15) for calculating geometric discord is quite questionable. This bound was derived in Ref.~\cite{LuoFuPRA10}. However, it is not so sharp and to improve it, the sharpest possible bound of geometric discord has been given in Refs. \cite{RanaParasharPRA12,HassanETALPRA12}. Indeed, the sharpness has been analytically proven in Ref.~\cite{HassanETALPRA12} and has been illustrated with several examples (notably all in $3\otimes 3$).

 Luckily, for this particular class of states, the bounds of Ref.~\cite{LuoFuPRA10} and Ref.~\cite{RanaParasharPRA12,HassanETALPRA12} coincide. However, due to the absence of condition for exactness of the bound in Ref.~\cite{YanLiuCheePRA13}, the authors are unable to judge the saturation of the bound. Using the optimal measurement suggested in Ref.~\cite{RanaParasharPRA12} (or Ref.~\cite{HassanETALPRA12}), it could be easily seen  (as has been shown in Ref.~\cite{OursarXivv1}) that the bound is exact for $\alpha\notin[\frac{5-\sqrt{5}}{2}, \frac{5+\sqrt{5}}{2}]$, which leads to the analytic value $\alpha=\frac{5\pm\sqrt{5}}{2}$ where \textit{ sudden change of discord} occurs. But within this range of $\alpha$, there is no known measurement agreeing with the bound and hence exact value of discord is unknown---all that is known is only a finer upper and lower bound (see Eq. (13) of \cite{OursarXivv1}).

We also wish to point out that the claim ``sudden change at transition point from bound to free entanglement..." need not be over emphasized. The simple fact is that if a state $\rho$ depends on a parameter $\alpha$, then the discord $\D(\rho)$ would obviously be a function of $\alpha$, and in particular this function may be piece-wise as well. 

\emph{Note:} Ref.~\cite{OursarXivv1} was available as \href{http://arxiv.org/abs/1207.1652v1}{arxive 1207.1652v1} which contains this result as a subsection.

\end{document}